# Cross-disciplinary learning:
# A framework for assessing application of concepts across STEM disciplines

Emily Borda
*Department of Chemistry and Science, Math, and Technology Education Program*
*Western Washington University*
*Bellingham, WA*

Todd Haskell
*Department of Psychology*
*Western Washington University*
*Bellingham, WA*

Andrew Boudreaux
*Department of Physics and Astronomy*
*Western Washington University*
*Bellingham, WA*

14 Dec 2020

**Abstract**

We propose *cross-disciplinary learning* as a construct which can guide instruction and assessment in programs that feature sequential learning across multiple STEM disciplines. Cross-disciplinary learning combines insights from interdisciplinary learning, transfer, and resources frameworks and highlights the processes of resource activation, transformation, and integration to support sensemaking in a novel disciplinary context by drawing on knowledge from other, prerequisite disciplines. We describe two measurement approaches based on this construct: A paired multiple choice instrument set to measure the extent of cross-disciplinary learning, and a think-aloud interview approach to provide insights into which resources are activated, and how they are used, in making sense of an unfamiliar phenomenon. We offer implications for program and course assessment.

**Introduction**

Amara is beginning her second term at university. She hopes to become an engineer and has thrown herself enthusiastically into her coursework. Despite this, she struggled in first-semester physics and chemistry. In her second chemistry course, she is learning about exothermic and endothermic reactions. When the class discusses the combustion of methane as an exothermic reaction, Amara is puzzled that energy seems to come from nowhere. She remembers the Law of Conservation of Energy from physics and thinks, "The energy must come from somewhere, but from where?" Her professor says something about energy being stored in the methane, which makes Amara think of potential energy, another idea she studied in physics. She recalls an example of a skateboarder riding down a hill, in which potential energy was transformed into kinetic energy. She wonders if potential energy is transforming as methane burns, but isn't sure what potential energy is for methane or what other type of energy it is transformed into.

Miguel is in the same second-term chemistry course. An aspiring biochemist, he studied hard during his first term and earned top grades in both physics and chemistry. Now, during the class discussion of methane combustion, he writes down notes and definitions, highlights words like enthalpy, and memorizes heuristics such as *exothermic = exit* and *endothermic = enter*. He launches into learning these new facts. "Energy again," he thinks. "Enthalpy must be a new type of energy for chemistry. Glad we're not talking about that potential energy stuff anymore."

Many instructors can likely think of Amaras and Miguels at their own institutions. While Amara exhibits an approach to learning we hope to instill in our students, we suspect Miguel's approach is more likely to lead to academic success. We aspire for students to seek connections between disciplines, but recognize that common academic structures and assessment practices, including those at our own institution, may reward students who focus narrowly on the content and problem-solving skills associated with a single course.

Here we propose the construct *cross-disciplinary learning*, drawing on several well-established theoretical frameworks, and describe two associated assessment approaches, one quantitative and one qualitative. We present these approaches as tools for making cross-disciplinary learning visible, so that it can be promoted at the classroom, departmental, and inter-departmental levels. In future papers we intend to detail the development and validation of the instruments and protocols and present a comprehensive analysis of data they have generated. Here we limit ourselves to describing our approach, in the hope of stimulating discussion about how to see, reward, and further develop habits of mind like those of Amara.

## Significance

The sequenced structure of many undergraduate STEM programs is consistent with two underlying assumptions about learning: first, that concepts and skills acquired in earlier courses serve as building blocks for mastering content in later courses, and second, that students will productively combine them to support their learning in new disciplinary contexts. Shared concepts can be used to bridge knowledge from different science disciplines. For example, the general rule that the potential energy of a system of attracting objects increases with object separation can be used to make sense of why a ball slows down when thrown upward and why energy input is needed to break a chemical bond. While students could reason by analogy directly from one situation to the other (Gentner, 1989), knowledge of shared concepts will help them generalize to other situations involving attracting objects, regardless of the discipline.

Our work has focused on energy, a fundamental framework for building explanations in all science disciplines (Eisenkraft et al., 2014). Its importance is highlighted as both a core idea and a cross-cutting concept in the Next Generation Science Standards (NGSS Lead States, 2013). The siloed nature, distinct foci, and technical vocabulary of the disciplines, however, can obscure the universal applicability of energy (Eisenkraft et al., 2014). Furthermore, as illustrated in the opening vignette, course and program assessments may not be sensitive to student learning of this kind of interconnected knowledge.

Though not the norm, course and program designs that explicitly promote cross-disciplinary learning do exist. Indeed, the insights driving our approach originated from our work in such a course sequence. Our institution offers a set of coordinated science content courses for preservice elementary and middle school teachers, with one course each in physics, geology, biology, and chemistry. The curricula (Authors; Deborah A. Donovan et al., 2008; Goldberg,

Robinson, & Otero, 2005; Goldberg et al., 2016) all focus on the flow of matter and energy, use similar vocabulary, representations, and modeling tools surrounding energy concepts, and employ a common, constructivist-based pedagogy (Baviskar, Hartle, & Whitney, 2009; Colburn, 2000). The coherent nature of this series has made it a valuable laboratory for exploring students' cross-disciplinary learning of energy. Common themes and pedagogies provide opportunities for students to make connections across disciplines, and for those connections to be more visible than they might be in traditional science courses (Authors). We hoped this course series would allow us to see, characterize, and measure cross-disciplinary learning. For example, students develop and use *energy diagrams* – a specific representation – throughout the series, and use common vocabulary around energy forms and transfer types. As students encounter new energy forms and transfer mechanisms, they reason and explain using the familiar energy diagram representation – allowing us to evaluate whether and characterize how students apply foundational energy concepts in new disciplinary contexts.

Although our work in this instructional setting has played an important role in the development of the assessment approaches described in this paper, we recognize that a siloed structure remains the norm in STEM education, including at our own university. We believe that the assessment tools we have developed for characterizing cross-disciplinary learning are appropriate for use in traditional, more siloed course sequences. It is perhaps in that context where making cross-disciplinary learning visible is most valuable.

**Conceptualizing and measuring cross-disciplinary learning**

In developing and teaching the course series mentioned above, we increasingly felt the need to gauge how well the courses were helping students apply energy ideas across different disciplinary contexts. Achieving this goal required a theoretically-grounded construct that foregrounds the process of making connections. Below we draw on existing theoretical approaches to help define cross-disciplinary learning. We then apply this construct to describe novel approaches to assessment.

**Defining Cross-Disciplinary Learning**

The vignette featuring Amara illustrates the kind of science learning we hope our students will achieve. In this vision, concepts and skills acquired in foundational courses are combined to support learning in later courses, including in novel disciplinary contexts. We refer to this process as "cross-disciplinary learning," and provide a model, featuring Amara's reasoning, in Figure 1. Encountering a learning challenge in a novel disciplinary context triggers *activation* of knowledge developed in previous coursework, such as ideas about potential energy and energy conservation from Amara's physics course, as well as knowledge from current coursework, such as the observation of increasing temperature during methane combustion. Some activated knowledge elements are *transformed* as Amara combines them. For example, the idea that potential energy transforms into kinetic energy, originally learned in the physics context, is recast in a more general form, which can be more readily combined with the knowledge from chemistry that heat is released when methane combusts. The *integration* of these energy ideas then supports the development of a new insight, that potential energy associated with methane transforms during combustion. Overall, cross-disciplinary learning involves active synthesis, possible through the (perhaps implicit) recognition that some ideas (*e.g.,* energy cannot be created or destroyed) apply to situations with widely disparate surface features (*e.g.,* skateboarding and methane combustion).

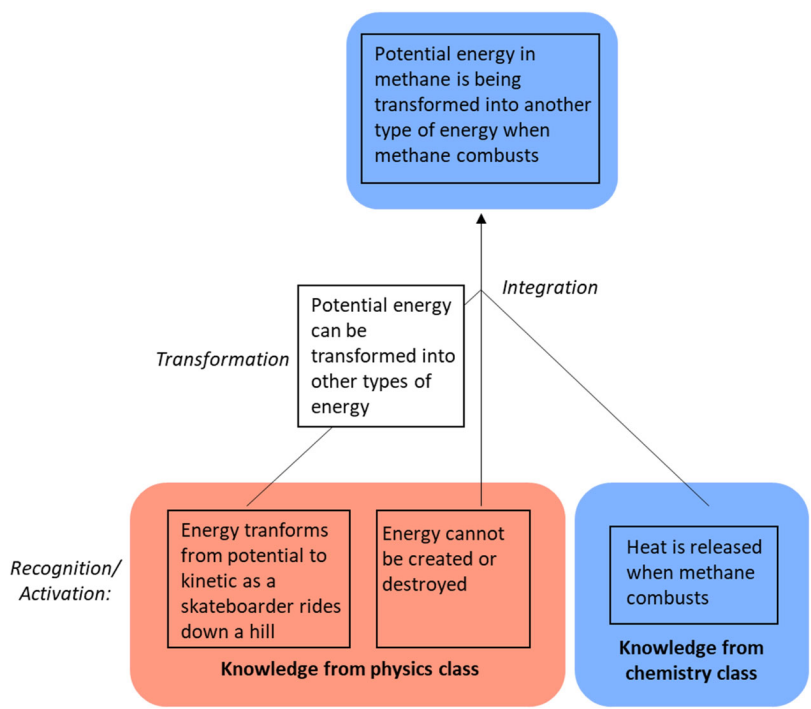

**Figure 1.** A model for cross-disciplinary learning.

To support and further elucidate this model for cross-disciplinary learning, we make connections with existing relevant theoretical frameworks. These include *interdisciplinary learning, transfer,* and *resources.* We describe how each of these relates to cross-disciplinary learning, and illustrate how this novel construct can provide practical guidance for developing assessments that will be sensitive to the type of sensemaking exhibited by Amara.

**Interdisciplinary learning.** Shen, Sung, and Zhang (2015) propose an Interdisciplinary Reasoning and Communication (IRC) framework for how individuals leverage knowledge from multiple disciplinary contexts to solve a complex problem. Two features of this framework are *transformation* and *integration*, which play a central role in our model of cross-disciplinary learning. Transformation involves changing the form of a knowledge element to enhance its applicability to a new context, while integration involves combining different elements.

Examples of transformation we have observed in student reasoning include generalizing from gravitational interactions to other types of attractive interactions (e.g. magnetic, electrostatic) when reasoning about potential energy changes, and extending the concept of kinetic energy from the context of a single macroscopic object to the context of the small particles that make up a bulk material when reasoning about temperature changes in a chemical system. Examples of integration include connecting the concept of potential energy to the idea of ATP as an energy carrier to make sense of energy changes in living organisms, and combining a small particle model of matter with the notion of heat transfer to develop particle collisions as a mechanism for temperature changes in thermal interactions.

Interdisciplinary learning frameworks foreground a *problem* or *context* and seek to identify the different ideas students combine in finding solutions (Shen, Liu, & Sung, 2014). Such frameworks are especially applicable for integrated science or problem-based learning curricula. We adopt a constructivist lens (Colburn, 2000; Piaget, 1978; Vygotsky, 1978) to match our focus on more sequential learning. Our focus on sequential learning is somewhat different from the problem-based approaches described in some interdisciplinary learning frameworks. Through a constructivist lens, learning is an active, social process learners engage in to develop ideas based on experience and evidence, rather than a passive assimilation of knowledge. We therefore foreground the *new ideas and knowledge* developed as a part of this active process as the overall goal of cross-disciplinary learning.

**Transfer.** Transfer frameworks are relevant to cross-disciplinary learning in part through their shared focus on sequential learning. Two specific kinds of transfer are relevant to our conception of cross-disciplinary learning: direct transfer and Preparation for Future Learning (PFL). Direct, or classic, transfer research has examined the extent to which students apply learned concepts, reasoning, or problem solving skills directly to a new target domain (e.g. Gick & Holyoak, 1983). Traditional assessment methods often fail to detect direct transfer when there are many differences between the learning and application contexts – so called "far transfer" (Barnett & Ceci, 2002). Dreyfus et al. (2014), for example, describe a student struggling to adapt the idea of potential energy stored in gravitational fields to potential energy associated with chemical bonds to explain energy changes during chemical reactions. The gravitational field conception does not support the use of potential energy in an explanation for chemical reactions, and thus cannot be transferred directly to leverage this explanation. As this example illustrates, cross-disciplinary learning often requires transformation of concepts for productive connections to be made. Thus, while we see direct transfer as an important theoretical idea to draw on, for our purposes it was important to foreground the integrative and transformative processes involved in cross-disciplinary learning.

In contrast to direct transfer, in which a learned concept or skill is applied "as-is" to a novel challenge, Preparation for Future Learning (PFL) frames prior learning experiences as preparation for an active process of constructing knowledge in the new context (Bransford & Schwartz, 1999; Schwartz, Bransford, & Sears, 2005). This view has clear overlap with cross-disciplinary learning: Knowledge developed in a prerequisite course provides students with tools for reasoning and learning in a new context. For example, familiarity with energy conservation from physics led Amara to ask questions about what forms of energy are changing and how potential energy may be involved when learning about combustion in chemistry.

**Resources.** To transform and integrate knowledge from different disciplines, Amara must first activate relevant knowledge learned in those contexts. Conceptual resource frameworks describe small-grained, existing bits of knowledge that are marshalled to form larger ideas or to explain a phenomenon (Benedikt, Virginia, & Michael, 2013; Richards, Jones, & Etkina, 2018). Resources related to energy include associating an energy form with a physical indicator, and accounting for energy quantities before and after a process (Sabo, Goodhew, & Robertson, 2016). Resources function as building blocks of knowledge that are activated in context-specific ways. Because learning is defined in terms of the activation of existing cognitive entities, learning and transfer are not strongly distinguished in a resources framework (Hammer, Elby, Scherr, & Redish, 2005).

Combining these frameworks, we define cross-disciplinary learning as a process of sequential learning in which students activate, transform, and integrate knowledge from different disciplines, combining previous learning with new learning to construct new knowledge within a specific discipline. We propose cross-disciplinary learning not as an alternative to these theoretical frameworks, but rather as a combination of their elements that captures the learning that traditional STEM programs seem to expect students to engage in. The construct of cross-disciplinary learning is a useful guide when evaluating where that kind of learning is occurring, and where it is not.

**Measuring cross-disciplinary learning**

Traditional, siloed science instruction relegates most or all responsibility for making connections between disciplines to the student. By offering approaches to measuring cross-disciplinary learning, we hope to support instructors in actively promoting it for their students, thus shifting to a model of shared responsibility. The approaches to assessment we describe below could be used to generate formative feedback, at a classroom level, that could help faculty adjust instruction in real time. Assessments could also be used for more summative, program-level evaluation, which could spur collaborative work toward more coherent, integrated curricula. We offer both quantitative and qualitative approaches. While quantitative assessment provides insight into the extent to which different conceptual resources are activated, a qualitative approach can reveal what resources are marshalled and how they are transformed and combined.

**Quantitative measurement of cross-disciplinary learning**. Quantitative instruments can be used at the classroom level at the beginning of a term to help guide instruction, at the end of the term to show growth, or at the program level to motivate curricular changes. Several instruments exist to measure students' understanding of energy in multiple science disciplines (Herrmann-Abell & DeBoer, 2014; Opitz, Neumann, Bernholt, & Harms, 2017; Park & Liu, 2016). To assess cross-disciplinary learning, we sought an approach that would be sensitive to students' ability to use energy ideas to "figure things out" in an unfamiliar context, while backgrounding their ability to apply those concepts within each discipline. We also wanted assessments to be practical to use to inform teaching or program improvement. These considerations led us to develop a pair of instruments that assesses a single energy idea – the conservation principle – in two disciplinary contexts: one that would be familiar to students, and one more novel. We chose physics as the familiar context because the physics course is a prerequisite in the course series that has guided this work, and chemistry as the unfamiliar context, because most of our students take chemistry last in the sequence. Physics commonly serves as a prerequisite to many STEM programs in our, and other, universities, as well. Finally, students' difficulty translating energy ideas from physics to chemistry is well documented (Becker & Cooper, 2014; Dreyfus et al., 2014; Lindsey, Nagel, & Savani, 2019; Nagel & Lindsey, 2015), establishing a need for tools to support cross-disciplinary teaching and learning in this area.

As an example, Figure 2 shows one item from each instrument. Both questions involve the energy conservation principle, but the first question is in a physics context, a child throwing a ball, and the second is in a chemistry context, an electron moving in a Bohr atom. (Note that we are less concerned about the sophistication or currency of this model of the atom than we are about the students' ability to reason with it). Items from the instrument featuring the familiar disciplinary context "screen" for the activation of a conceptual resource. The physics question, for example, requires the idea that the kinetic and gravitational potential energy associated with projectile motion will obey "accounting rules." Items in the less familiar context require a student to transform this conceptual resource and integrate it with knowledge specific to the new context. For example, on the chemistry question the accounting resource must be integrated with knowledge that in the Bohr atom, an electron can "jump" to a higher shell, which corresponds with greater separation from the attracting proton and therefore higher potential energy.

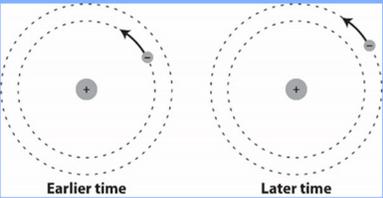

**Figure 2.** An example of an item pair for assessing cross-disciplinary learning.

For performance on this instrument pair to depend strongly on student ability to reason across disciplinary boundaries, we wanted students to have sufficient information about the novel context to reason *with*. The chemistry items are grouped into four contexts: an electron moving in a Bohr atom, two ions forming a bond, phase changes in a pure substance, and chemical reactions. Each context is preceded with a narrative that introduces energy forms and indicators pertaining to it. These narratives are followed by unscored reading comprehension questions, which give students feedback to help them refine their ability to use these new resources. The narratives and reading comprehension questions are built into the assessment instrument itself, in an effort to foreground student reasoning (e.g. transformation and integration of concepts and knowledge), rather than technical understanding of new vocabulary, in the measurement of cross-disciplinary learning. Figure 3 shows the narrative and reading comprehension questions that precede the Bohr model items.

Some instructors may have concerns about providing chemistry content as part of the assessment. However, this measurement approach is not intended to measure student understanding *in* either of the disciplines, but rather students' skill in leveraging their knowledge from one discipline to figure out new knowledge in the other (we administered the two instruments *after* physics instruction but *before* any relevant chemistry instruction). These instruments do not replace concept inventories, which can measure students' understanding of relevant concepts as instruction proceeds.

> At the center of an atom is the nucleus, which contains particles with positive electric charge called protons. The nucleus is surrounded by negatively charged particles called electrons. In this model of the atom, electrons possess *kinetic energy* and orbit the nucleus at some speed. The greater the speed of an electron, the more kinetic energy it has.
>
> Since protons and electrons have opposite electric charge, they are attracted to one other. Due to this attraction, electron-nucleus systems have *Electrostatic Potential Energy* (EPE). When an electron moves farther away from the nucleus, the EPE increases. This is similar to the gravitational interaction between two objects: in each case, potential energy increases as two attracting objects become farther apart.
>
> The total energy of an atom consists of the sum of the kinetic and potential energy of the atom's nucleus and electrons. The total energy of an atom can change over time as the atom interacts with its environment. In addition, electrons can change their distance from the nucleus by "jumping" to an orbital path, or "shell," that is farther away from the nucleus (a "higher" shell) or closer to the nucleus (a "lower" shell). A simple model of a hydrogen atom, which consists of one proton and one electron, is shown at right.
>
> 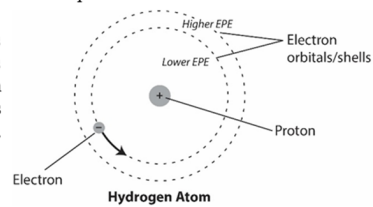
>
> 1. As an electron moves away from the nucleus of an atom, the atom-nucleus system:
> a) Decreases in EPE
> b) Increases in EPE
> c) Retains the same amount of EPE
>
> 2. An atom has 5 units of kinetic energy and 6 units of potential energy. What is its total energy?
> a) -1 unit
> b) 0 units
> c) 1 unit
> d) 11 units
> e) 30 units

**Figure 3.** An example of a teaching narrative followed by two reading comprehension questions.

The paired instruments share a common requirement for success: the activation of conceptual resources for energy conservation. The chemistry instrument provides additional information specific to the chemistry context, through the narratives, and adds the requirement that students integrate the energy conservation resources learned in physics with the newly introduced conceptual elements from chemistry. In order to describe a student's ability to reason across disciplinary contexts, an instructor can compare their relative performance on the physics and chemistry instruments. Strong relative performance on the chemistry instrument suggests the student is able to combine already learned knowledge, as measured by the physics instrument, with the new information provided about the unfamiliar context. Relatively weaker performance on the chemistry measure suggests more compartmentalized learning. This allows instructors to disentangle how much the student learned about physics from their ability to apply whatever they learned to a novel context, and could assess course designs that explicitly teach cross-disciplinary learning skills. Since the chemistry instrument is organized into several sub-contexts, the instructor can also gain specific understandings of where students are succeeding and struggling with cross-disciplinary learning and modify instruction as needed.

During administration, students complete the chemistry items before completing the physics items. This ensures the physics instrument does not itself act as a treatment, by cueing students to apply particular energy ideas, but rather measures the extent to which students understood the energy ideas they were trying to apply. A future manuscript will present the instrument pair (available upon request) in detail, describe its validation, and share results. The paired instrument approach can be used for program-level assessment, while instructors can use a paired question approach, perhaps asking for written follow-up explanations, on clicker questions, exit slips, quizzes, and exams, to formatively and summatively assess students' cross-disciplinary learning in their courses. A paired question approach can also support discussions about, and seed expectations for, cross-disciplinary learning.

**Qualitative measurement of cross-disciplinary learning**. Qualitative assessment can provide rich depictions of how conceptual resources are combined and transformed as a student formulates an explanation in a novel disciplinary context. A qualitative approach is particularly well suited for formative assessment, since knowledge of how students use conceptual resources can help the instructor guide students to connect new material to previous learning. In parallel with the quantitative approach, we developed a set of interview protocols for students in the science education sequence. The protocols are designed for use with students who have completed the prerequisite physics course and just started either the geology, biology, or chemistry course. These students had developed energy concepts in a physics context, but had not yet started to systematically apply those concepts in the subsequent discipline.

To elicit student thinking, we used a Think-Aloud Interview (TAI) approach (Bowen, 1994) and asked students to construct a scientific explanation for an observed phenomenon. Each protocol involved three such scenarios: two for the less familiar disciplinary context (either geology, biology, or chemistry), and one for the more familiar physics context. The same energy concepts are relevant for all three scenarios. As in our quantitative assessment, we present the physics scenario at the end of the interview, so the interview itself would not cue students to apply relevant concepts in the novel discipline. For consistency with the existing discussion, we have shown our chemistry protocol as an example in Figure 4.

The interview protocols are semi-structured, and each require 30-50 minutes. Students were presented with a sequence of questions from the protocol. The interviewer deployed follow-up questions as needed to further draw out student thinking. Interviews were recorded using a Smart Pen. Participants were recruited from science education courses and compensated with gift cards. All provided informed consent.

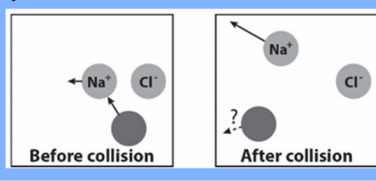

**Figure 4.** Sample interview protocol used for qualitative assessment of cross-disciplinary learning.

Data collection and analysis thus far suggests this approach is useful for exploring what resources students activate and in what combinations, as well as how the resources are used in making sense of novel scenarios. As with the quantitative instruments, a detailed description of this work will be the subject of a future manuscript, and the protocols are available upon request from the authors. We believe the basic structure of the interviews, involving open-ended questions that pair two contexts using parallel ideas, could be modified as exit slips or homework for course-level formative assessment. The interviews could also be conducted with a sample of majors or alumni as part of a summative, program assessment plan.

## Conclusion

We have proposed the construct of cross-disciplinary learning and have begun to operationalize it by outlining frameworks for measurement. We have focused our discussion on energy in physics and chemistry, but the approach could be adapted to other cross-cutting concepts and disciplinary contexts as well. Paired multiple choice items and cross-disciplinary think-aloud interview tasks can foreground resource activation and integration, while reducing the probability that context-specific, procedural or declarative knowledge (or its absence) will solely determine the quality of student responses.

We do not suggest discipline-specific learning outcomes and assessments be replaced or abandoned. Rather, we argue assessments that focus narrowly on a single disciplinary context can be supplemented with approaches that explicitly target ways of making connections across disciplines. Such assessments could reward and encourage the type of thinking exhibited by Amara and motivate and assess the creation of curricula and pedagogies that would support students like Miguel in moving toward more integrative habits of thinking.

Developing assessments and pedagogy focused on cross-disciplinary learning requires collaboration between disciplinary units to determine learning outcomes, frame them with common vocabulary, and develop curricular coherence. This may be especially true in introductory coursework, in which cross-cutting ideas come up frequently but may be obscured from students by discipline-specific vocabulary and representations. Perhaps most of all, prioritizing cross-disciplinary learning challenges us to think of ourselves as STEM educators first and disciplinary experts second. We hope this work supports conversations about how to envision coherent teaching and learning across our STEM programs.

**Acknowledgements**

This work was supported by the National Science Foundation (DUE-1612055, DUE-1612251). Any opinions, findings, conclusions or recommendations expressed in this material are those of the authors and do not necessarily reflect the views of the National Science Foundation. Remaining acknowledgments are removed for blinding.

**References**

Barnett, S. M., & Ceci, S. J. (2002). When and where do we apply what we learn? A taxonomy for far transfer. *Psychological Bulletin, 128*(4), 612-637. doi:10.1037//0033-2909.128.4.612

Baviskar, S. N., Hartle, R. T., & Whitney, T. (2009). Essential criteria to characterize constructivist teaching: Derived from a review of the literature and applied to five constructivist-teaching method articles. *International Journal of Science Education, 31*(4), 541-550.

Becker, N. M., & Cooper, M. M. (2014). College chemistry students' understanding of potential energy in the context of atomic-molecular interactions. *Journal of Research in Science Teaching, 51*(6), 789-808. doi:10.1002/tea.21159

Benedikt, W. H., Virginia, J. F., & Michael, C. W. (2013). Productive resources in students' ideas about energy: An alternative analysis of Watts' original interview transcripts. *Physical Review Special Topics. Physics Education Research, 9*(2), 023101. doi:10.1103/PhysRevSTPER.9.023101

Bowen, C. W. (1994). Think-aloud methods in chemistry education: Understanding student thinking. *Journal of Chemical Education, 71*(3), 184. doi:10.1021/ed071p184

Bransford, J. D., & Schwartz, D. L. (1999). Rethinking transfer: A simple proposal with multiple implications. *Review of Research in Education, 24 1999, 24*, 61-100. doi:10.3102/0091732x024001061

Colburn, A. (2000). Constructivism: Science education's "Grand Unifying Theory". *The Clearing House: A Journal of Educational Strategies, Issues and Ideas, 74*(1), 9-12. doi:10.1080/00098655.2000.11478630

Donovan, D. A., Atkins, L. J., Salter, I. Y., Gallagher, D. J., Kratz, R. F., Rousseau, J. V., & Nelson, G. D. (2013). Advantages and challenges of using physics curricula as a model for reforming an undergraduate biology course. *Cbe-Life Sciences Education, 12*(2), 215-229. doi:10.1187/cbe.12-08-0134

Donovan, D. A., Rousseau, J., Salter, I., Atkins, L., Acevedo-Gutierrez, A., Kratz, R., . . . Pape-Lindstrom, P. (2008). *Life Science and Everyday Thinking*. Greenwich, CT: Activate Learning.

Dreyfus, B. W., Gouvea, J., Geller, B. D., Sawtelle, V., Turpen, C., & Redish, E. F. (2014). Chemical energy in an introductory physics course for the life sciences. *American Journal of Physics, 82*(5), 403-411. doi:10.1119/1.4870391

Eisenkraft, A., Nordine, J. C., Chen, R. F., Fortus, D., Krajcik, J., Neumann, K., & Scheff, A. (2014). Why focus on energy instruction? In R. F. Chen, A. Eisenkraft, D. Fortus, J. Krajcik, K. Neumann, J. C. Nordine, & A. Scheff (Eds.), *Teaching and Learning of Energy in K-12 Education* (pp. 1-5): Springer International Publishing.

Gentner, D. (1989). The mechanisms of analogical learning. In A. Ortony & S. Vosniadou (Eds.), *Similarity and Analogical Reasoning* (pp. 199-241). Cambridge: Cambridge University Press.

Gick, M. L., & Holyoak, K. J. (1983). Schema induction and analogical transfer. *Cognitive Psychology, 15*(1), 1-38. doi:10.1016/0010-0285(83)90002-6

Goldberg, F., Robinson, S., & Otero, V. (2005). *Physics and Everyday Thinking*. Armonk, NY: It's About Time: Herff Jones Education Division.

Goldberg, F., Robinson, S., Price, E., Harlow, D., Andrew, J., & McKean, M. (2016). *Next Generation Physics and Everyday Thinking*. Greenwich, CT: Activate Learning.

Hammer, D., Elby, A., Scherr, R. E., & Redish, E. F. (2005). Resources, framing, and transfer. In J. P. Mestre (Ed.), *Transfer of Learning: Research and Perspectives*. United States: Information Age Publishing.

Herrmann-Abell, C. F., & DeBoer, G. E. (2014). Developing and using distractor-driven multiple-choice assessments aligned to ideas about energy forms, transformation, transfer, and conservation. In R. F. Chen, A. Eisenkraft, D. Fortus, J. Krajcik, K. Neumann, J. C. Nordine, & A. Scheff (Eds.), *Teaching and Learning of Energy in K-12 Education* (pp. 103-133): Springer International Publishing.

Lindsey, B. A., Nagel, M. L., & Savani, B. N. (2019). Leveraging understanding of energy from physics to overcome unproductive intuitions in chemistry. *Physical review. Physics education research, 15*(1). doi:10.1103/PhysRevPhysEducRes.15.010120

Nagel, M. L., & Lindsey, B. A. (2015). Student use of energy concepts from physics in chemistry courses. *Chemistry Education Research and Practice, 16*(1), 67-81. doi:10.1039/c4rp00184b

NGSS Lead States. (2013). Next Generation Science Standards: For states, by states. In Washington, D.C.: National Academies Press.

Opitz, S. T., Neumann, K., Bernholt, S., & Harms, U. (2017). How do students understand energy in biology, chemistry, and physics? Development and validation of an assessment instrument. *Eurasia journal of mathematics, science and technology education, 13*(7), 3019. doi:10.12973/eurasia.2017.00703a

Park, M., & Liu, X. (2016). Assessing understanding of the energy concept in different science disciplines. *Science Education, Advance Release*.

Piaget, J. (1978). *Success and Understanding*. Cambridge, MA: Harvard University Press.

Richards, A. J., Jones, D. C., & Etkina, E. (2018). How students combine resources to make conceptual breakthroughs. *Research in Science Education*. doi:10.1007/s11165-018-9725-8

Sabo, H. C., Goodhew, L. M., & Robertson, A. D. (2016). University student conceptual resources for understanding energy. *Phys. Rev. ST Phys. Educ. Res., 12*(010126), 010126-010121-010128.

Schwartz, D. L., Bransford, J. D., & Sears, D. (2005). Efficiency and innovation in transfer. In J. P. Mestre (Ed.), *Transfer of Learning: Research and Perspectives*. United States: Information Age Publishing.

Shen, J., Liu, O. L., & Sung, S. (2014). Designing interdisciplinary assessments in sciences for college students: An example on osmosis. *International Journal of Science Education, 36*(11), 1-21. doi:10.1080/09500693.2013.879224

Shen, J., Sung, S., & Zhang, D. (2015). Toward an analytic framework of Interdisciplinary Reasoning and Communication (IRC) Processes in Science. *International Journal of Science Education, 37*(17), 2809-2835. doi:10.1080/09500693.2015.1106026

Vygotsky, L. S. (1978). *Mind in society: The development of higher psychological processes*. Cambridge: Harvard University Press.